\documentclass[reprint,amsmath,amssymb,aps,prc,superscriptaddress]{revtex4}
\usepackage{lineno}
\usepackage{graphicx}
\usepackage{dcolumn}
\usepackage{bm}
\usepackage[utf8]{inputenc}
\usepackage{comment}
\usepackage{subcaption}
\usepackage{hyperref}
\usepackage{tikz}
\usepackage{array}
\usepackage{tabularx}
\usepackage{multirow}
\usepackage{cleveref}
\usepackage{booktabs}
\newcolumntype{C}{>{$}c<{$}}

\AtBeginDocument{%
  \heavyrulewidth=.08em
  \lightrulewidth=.05em
  \cmidrulewidth=.03em
  \belowrulesep=.65ex
  \belowbottomsep=0pt
  \aboverulesep=.4ex
  \abovetopsep=0pt
  \cmidrulesep=\doublerulesep
  \cmidrulekern=.5em
  \defaultaddspace=.5em
}

\usepackage{xcolor}
\setcitestyle{maxcitenames=1}
\newcolumntype{P}[1]{>{\centering\arraybackslash}p{#1}}
\newcolumntype{M}[1]{>{\centering\arraybackslash}m{#1}}
\usepackage{pdfpages}

\newcommand{\mean}[1]{\left\langle #1 \right\rangle}

\hypersetup{%
    colorlinks=true
    ,urlcolor=blue
    ,citecolor=blue
    ,linkcolor=blue
    }

\usepackage{caption}
\begin{document}

\title{Proton High-Order Cumulants in Au+Au Collisions at High Baryon Density from JAM with a Centrality-Independent Framework}

\author{Yongcong Xu}
\author{Zhaohui Wang}
\affiliation{Key Laboratory of Quark \& Lepton Physics (MOE) and Institute of Particle Physics, \\ Central China Normal University}
\author{Yu Zhang}
\affiliation{Guangxi Normal University}
\author{Xiaofeng Luo } \thanks{xfluo@ccnu.edu.cn} 
\affiliation{Key Laboratory of Quark \& Lepton Physics (MOE) and Institute of Particle Physics, \\ Central China Normal University}

\date{\today}
\begin{abstract}
The event-by-event higher-order cumulants of conserved quantities such as 
net-baryon, net-electric charge, and net-strangeness in heavy-ion collisions 
have been extensively utilized in experimental searches for the QCD critical point, 
notably in the RHIC-STAR experiment. In this study, we conduct a systematic analysis of higher-order cumulants of proton number distributions in Au+Au collisions at center-of-mass energies of $\sqrt{s_{\rm NN}} = 3.2$, $3.5$, $3.9$, and $4.5$~GeV using the JAM model.
We calculate cumulants, factorial cumulants, and their ratios 
using a novel method,  Centrality-Independent Genuine Cumulant Analysis fRamework (CIGAR),
which effectively eliminates initial volume fluctuations. 
We comprehensively compare the CIGAR method with the traditional Centrality Bin Width Correction (CBWC) method.
In addition, the effect of spectators on cumulant is systematically investigated. 
Our results provide a dynamic non-critical baseline in the high-baryon-density regime
which is crucial for QCD critical point searches in heavy-ion collisions.

\end{abstract}
\maketitle

\section{Introduction}
One of the primary objectives of heavy-ion collision experiments is 
to investigate the properties of the quark-gluon plasma (QGP) 
and the phase structure of strongly interacting matter, which is conventionally characterized by the QCD phase diagram 
in terms of temperature and baryon chemical potential~\cite{Bzdak:2019pkr,Braun-Munzinger:2026krf,Luo:2022mtp}. 
Recent theoretical studies predict the existence of a QCD critical point (CP) 
at the end of the first-order phase transition line~\cite{Clarke2024, PhysRevD.110.094006, PhysRevD.101.054032, PhysRevD.104.054022, GAO2021136584, sorensen2024,Cai:2024eqa,Zhu:2025gxo}. 
Precisely determining the location of this critical point is crucial for elucidating the QCD phase structure and is a central scientific goal of contemporary heavy-ion experimental programs~\cite{Luo2017}.

High-order cumulants of conserved charges, including net-electric charge, net-baryon number, and net-strangeness, 
have been identified as promising observables for experimentally searching the QCD critical point due to 
their sensitivity to the correlation length $\xi$ that diverges at the critical point~\cite{Stephanov:2008qz,ASAKAWA2016299, PhysRevC.96.024910}. Therefore, non-monotonic dependence on collision energy would be a characteristic signal of the QCD critical point in heavy-ion collisions~\cite{Li:2017ple,Li:2018ygx,Fu:2023lcm,Lu:2026ezr}.
Since 2010, the RHIC-STAR Collaboration has measured high-order cumulants of net-proton numbers (proxy for net-baryon) in Au+Au collisions across a broad energy range $\sqrt{s_{\rm NN}}=$ 3 - 200~GeV in both collider mode (7.7 - 200 GeV) and fixed-target mode (3 - 7.7 GeV). At higher energies ($\sqrt{s_{\rm NN}}=$ 7.7 - 200~GeV), a non-monotonic structure of net-proton $C_4/C_2$ ratio is observed, with a maximum deviation below non-CP baseline at 19.6 GeV~\cite{PhysRevLett.105.022302, PhysRevLett.112.032302, PhysRevLett.113.092301, PhysRevC.100.014902, PhysRevLett.126.092301, PhysRevC.104.024902, PhysRevLett.127.262301, PhysRevLett.130.082301, Chen2024, STAR:2025zdq}. 
Meanwhile, the published results at $\sqrt{s_{\rm NN}}=3.0$~GeV indicate that, if the critical point exists, it would lie above $3.0$~GeV~\cite{PhysRevLett.127.262301, PhysRevC.107.024908}. Thus, the experimental results and corresponding non-CP model baseline at low energy ( 3 - 7.7 GeV) would be very crucial to locate the QCD critical point. 

Previous investigations demonstrate that initial volume fluctuations significantly 
influence cumulant measurements derived using the Centrality Bin Width Correction (CBWC)~\cite{Luo_2013} method in the high-baryon-density regime. 
This experimental challenge can be mitigated by implementing a centrality-independent framework~\cite{WANG2025139984} for cumulant measurement. 
In this study, we implement this novel method to calculate higher-order proton cumulants in Au+Au collisions 
at $\sqrt{s_{\rm NN}}=3.2$--$4.5$~GeV using the JAM model. 
Our results are systematically compared with those obtained from conventional CBWC approaches. In addition, we investigate the impact of spectator nucleon contributions.

The paper is structured as follows. 
Sec.~\ref{jam_model} describes the JAM model, 
the cumulant method, and the centrality-independent framework. 
Sec.~\ref{analysis} presents results of higher-order cumulant of proton distributions. 
Finally, Sec.~\ref{summary} summarizes the main findings.

\section{The JAM Model and Methodology}\label{jam_model}

\subsection{JAM Model}
 
 The JAM model (Jet AA Microscopic Transport Model)~\cite{2019JAM, PhysRevC.61.024901} is 
a non-equilibrium hadronic transport framework developed to simulate heavy-ion collisions 
across a broad energy range from approximately 100A~MeV up to RHIC energies. 
It employs a cascade method to propagate hadrons and excited states along explicit trajectories, with an energy-dependent treatment of inelastic hadron–hadron collisions. This treatment covers resonance production at low energies, string excitation at intermediate energies, and hard parton–parton scattering at high energies.
Previous phenomenological investigations utilizing the JAM model have successfully computed multiple cumulants 
and systematically examined various effects on particle number fluctuations in heavy-ion collision systems~\cite{He2016_JAM, Yu_JAM_5, Chatterjee2021_JAM}.
JAM, as its successor, provides a microscopic description of nuclear collision dynamics 
spanning from the initial stage to the final hadronic gas phase.

In this study, we perform Au+Au collision simulations at center-of-mass energies of 
$\sqrt{s_{\rm NN}}=3.2$, $3.5$, $3.9$, and $4.5$~GeV using the standard cascade mode of the JAM model. 
The impact parameter $b$ is sampled from $0$ to $14$~fm, 
with pseudorapidity coverage matching the acceptance of RHIC-STAR detector running at Fixed-target mode. 
The statistics for each center-of-mass energy are summarized in Table~\ref{tab_energy}.

\begin{table}[]
\begin{tabular}{@{}ccccc@{}}
\toprule
$\sqrt{s_{\rm NN}}$ / GeV            & 3.2 & 3.5 & 3.9 & 4.5 \\ \midrule
Events (with spectator) / million    & 55  & 49  & 58  & 58  \\
Events (without spectator) / million & 44  & 48  & 53  & 61  \\ \bottomrule
\end{tabular}
\caption{Statistics of Au+Au collisions simulated in JAM model at $\sqrt{s_{\rm NN}}=3.2-4.5$~GeV.} 
\label{tab_energy}
\end{table}

\begin{figure}[htbp]
    \centering
    \includegraphics[scale=0.5]{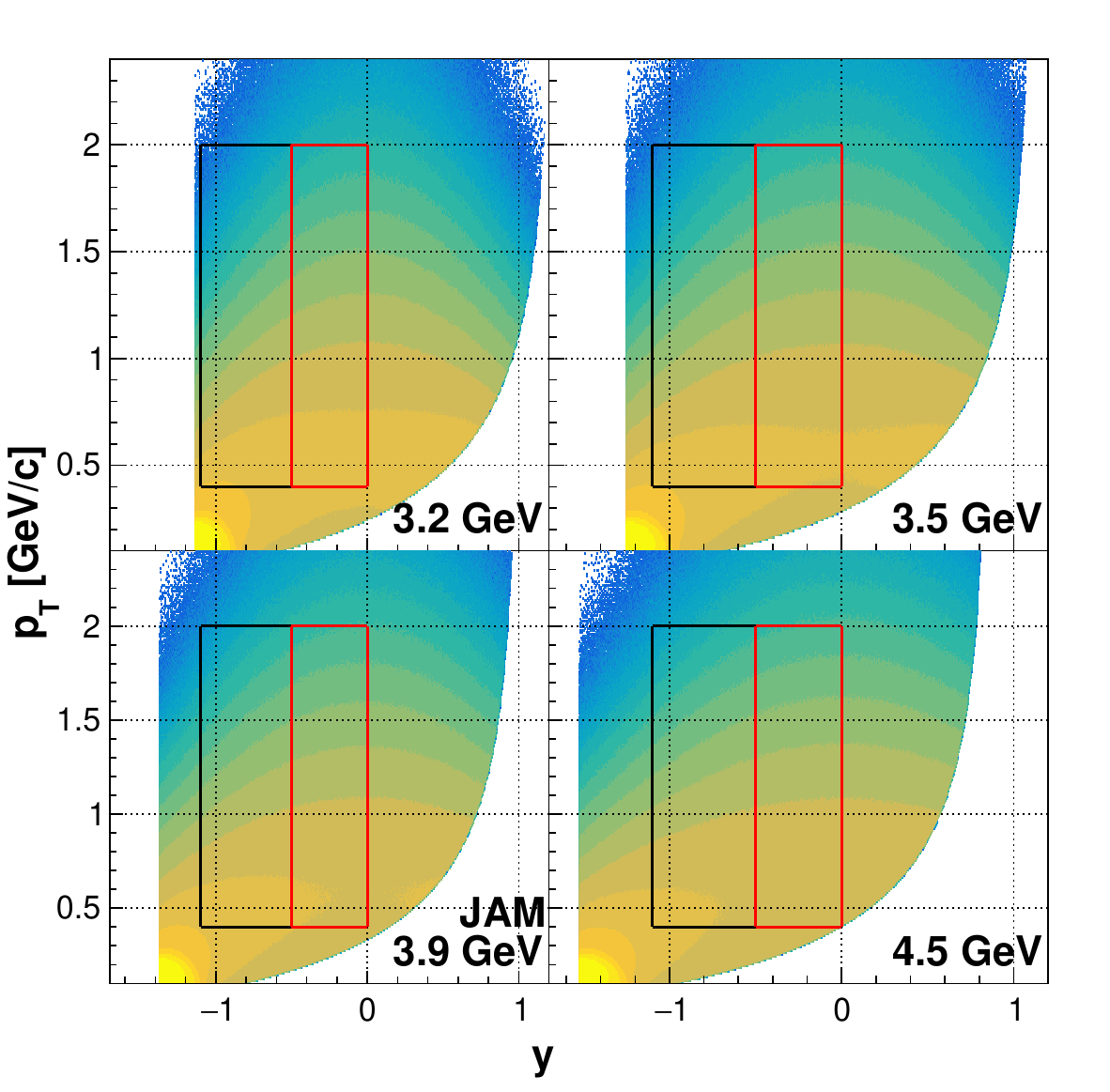}
    \caption{Proton acceptance in terms of transverse momentum $(p_{T})$ vs. proton rapidity$(y_{cm})$ in the center-of-mass frame in Au+Au collisions at $\sqrt{s_{\rm NN}} = 3.2$, $3.5$, $3.9$, and $4.5$~GeV from the JAM model. Protons are selected within a $p_{T}$ range of $0.4<p_{T}<2.0$~GeV/$c$, and rapidity range of $-1.1<y<0$.}
    \label{fig:AccWindow}
\end{figure}

\begin{figure}[htbp]
    \centering
    \includegraphics[scale=0.5]{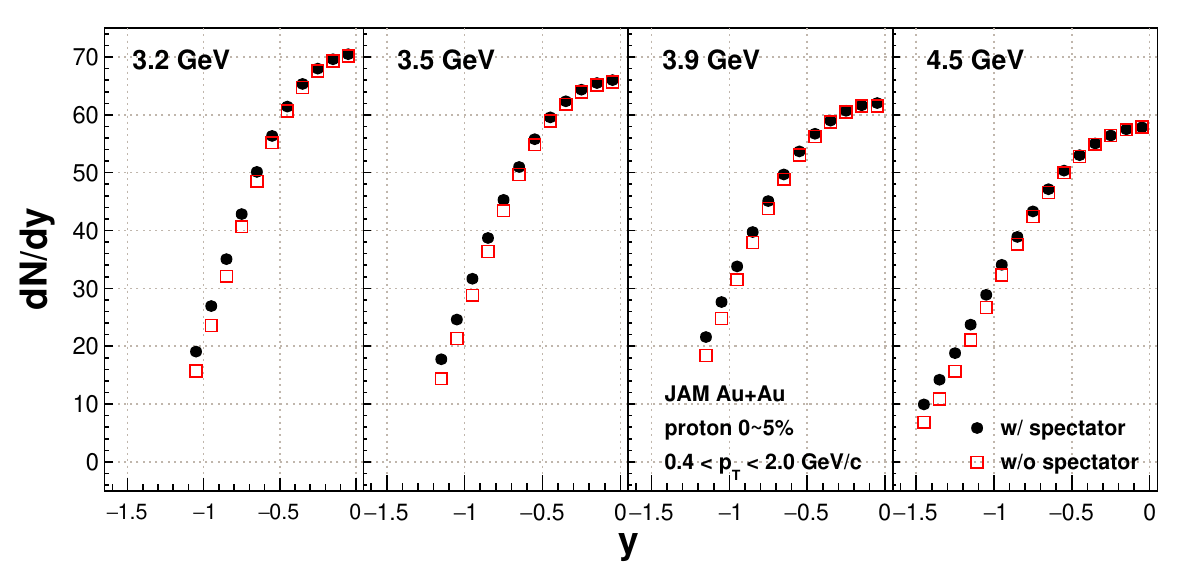}
    \caption{
        Proton dN/dy distributions within the $p_{T}$ range $0.4<p_{T}<2.0$~GeV/$c$ in 0-5\% central Au+Au collisions at $\sqrt{s_{\rm NN}}=3.2$, $3.5$, $3.9$, and $4.5$~GeV from JAM. Black solid circles represent the proton yield in the presence of spectator nucleons, while red open squares represent the yield without spectator nucleons. 
    }
    \label{fig:Dndy}
\end{figure}

Figure~\ref{fig:AccWindow} shows the proton acceptance in the transverse momentum ($p_{T}$) versus rapidity ($y_{cm}$) plane in the center-of-mass frame for Au+Au collisions at $\sqrt{s_{\rm NN}} = 3.2$, $3.5$, $3.9$, and $4.5$~GeV from the JAM model. The Au+Au simulations in JAM are performed in collider mode in the center-of-mass frame, and the final-state particles are subsequently boosted to the laboratory frame. A pseudorapidity cut of $-2.4 < \eta < 0$ is applied to match the STAR detector acceptance in fixed-target mode. The red boxes in each panel indicate the default acceptance window $-0.5 < y < 0$, while the black boxes mark the extended rapidity coverage up to $y = -1.1$, enabling a systematic study of the rapidity dependence of proton cumulants.

Figure~\ref{fig:Dndy} shows the proton dN/dy distributions within the $p_{T}$ range $0.4 < p_{T} < 2.0$~GeV/$c$ in 0--5\% central Au+Au collisions at $\sqrt{s_{\rm NN}} = 3.2$, $3.5$, $3.9$, and $4.5$~GeV from the JAM model. The black solid circles depict the proton yield in the presence of spectator nucleons, while the red open squares represent the yield without spectator contributions. At all four energies, the proton yield decreases monotonically from mid-rapidity ($y=0$) toward the beam rapidity. The yield is highest at $\sqrt{s_{\rm NN}} = 3.2$~GeV, with a value around 70 at mid-rapidity, and decreases systematically with increasing collision energy, dropping to approximately 60 at $\sqrt{s_{\rm NN}} = 4.5$~GeV. This trend reflects the weakening of baryon stopping at higher collision energies. Comparing the red open squares with the black solid circles, one can see the growing spectator effect at forward rapidities.

\subsection{Cumulant Definition}
The moments of a probability distribution, such as the mean, variance, skewness, and kurtosis characterize the shape of the distribution. The variance ($\sigma^2$) describes the width of the distribution, while the skewness and kurtosis quantify the asymmetry and peakedness around the mean, respectively. 

Let $P(N)$ be the probability distribution of a stochastic variable $N$. The moment generating function is written as 
    $M(t) = \langle e^{tN} \rangle$.
Then the $n$th-order moment $\langle N^n\rangle$ of $P(N)$ is defined as the coefficient of the Taylor expansion of $M(t)$ around $t = 0$.
\begin{equation}
    M(t) = \langle e^{tN} \rangle = 1 + t \langle N\rangle + \frac{t^2 \langle N^2 \rangle }{2!} + \frac{t^3 \langle N^3 \rangle }{3!} + ... + \frac{t^n \langle N^n \rangle }{n!}.
\end{equation}
Let $\delta N = N - \langle N \rangle$ be the deviation from the mean value. The $n$th-order central moment is then defined as 
$\mu_n = \langle (\delta N)^n\rangle, \quad n \geq 2$.
The cumulant generating function is written as 
$G(t) = \ln[M(t)] = \ln[\langle e^{tN} \rangle]$.
Similarly, the $n$th-order cumulant is defined as the coefficient of the Taylor expansion of $G(t)$.
\begin{equation}
    G(t) = \ln[\langle e^{tN} \rangle] = \sum^{\infty}_{n=1} C_{n} \frac{t^n}{n!} = C_{1} t + C_{2} \frac{t^2}{2!} + \cdots \label{eq:ExpendingSeries}
\end{equation}
In terms of central moments, cumulants $C_{n}$ are defined as 
\begin{equation}\label{eq:cumulant1}
	\begin{split}
		C_1 &= \mean{N}, \\
		C_2 &= \mean{(\delta N)^2} = \mu_2, \\
		C_3 &= \mean{(\delta N)^3} = \mu_3, \\
		C_4 &= \mean{(\delta N)^4} - 3\mean{(\delta N)^2}^2 = \mu_4 - 3\mu_2^2,\\
		C_{n} &= \mu_{n} - \sum^{n-2}_{m=2}\binom{n-1}{m-1}C_{m}\mu_{n-m},\quad {n}>3.
	\end{split}
\end{equation}
The commonly used cumulants—mean, variance, skewness, and kurtosis—are defined as:
\begin{equation}
\begin{split}
    \text{mean} &: M = \langle N \rangle = C_{1},\\
    \text{variance} &: \sigma^{2} = \langle (\delta N)^2 \rangle = C_{2},\\
    \text{skewness} &: S = \frac{ \langle (\delta N)^3 \rangle }  {\sigma^{3}} = \frac{C_{3}}{C_{2}^{\frac{3}{2}}},\\
    \text{kurtosis} &: \kappa = \frac{ \langle (\delta N)^4 \rangle } {\sigma^{4}} - 3 = \frac{C_{4}}{C_{2}^{2}}.\\
\end{split}
\label{eq:cumulantExplain}
\end{equation}
Cumulant ratios such as $C_{2}/C_{1}$, $C_{3}/C_{1}$, and $C_{4}/C_{2}$ are constructed to cancel volume dependence of leading order when comparing to theory:
\begin{equation}
    \frac{C_{2}}{C_{1}}=\frac{\sigma^{2}}{M},\quad \frac{C_{3}}{C_{1}}=\frac{S\sigma^{3}}{\mu},\quad\frac{C_{4}}{C_{2}}=\kappa\sigma^{2}.\label{eq:cumRatioCombination}
\end{equation}

The factorial cumulants are derived from the ordinary cumulants through the following relations:
\begin{equation}
\begin{split}
    \kappa_{1} &= C_{1} = \langle N \rangle ,\\
    \kappa_{2} &= -C_{1}+C_{2},\\
    \kappa_{3} &= 2C_{1}-3C_{2}+C_{3},\\
    \kappa_{4} &= -6C_{1}+11C_{2}-6C_{3}+C_{4}.\\
\end{split}
\label{eq:kappaEquation}
\end{equation}
Factorial cumulants offer the distinct advantage of directly isolating genuine $n$-particle correlations, as they vanish in the absence of such correlations.

\subsection{Centrality-Independent Genuine Cumulant Analysis Framework (CIGAR)}
We briefly introduce the centrality-independent analysis framework~\cite{WANG2025139984} in this section. This framework, termed CIGAR (Centrality-Independent Genuine Cumulant Analysis Framework) in the present study,
reconstructs the particle number distribution by optimizing Edgeworth expansion parameters through a hybrid approach 
combining differential evolution and Bayesian inference.
This methodology effectively addresses the initial volume fluctuation issue,
which is more pronounced in low-energy heavy-ion collisions than in high-energy ones.

Statistically, the probability density function of a distribution 
can be obtained through its cumulants using the Gram-Charlier series. 
As a viable solution to the moment problem, 
the Edgeworth expansion approximates a distribution $p(x)$ as a series expansion around the Gaussian distribution, 
incorporating higher-order cumulants to refine the shape characterization.
\begin{figure}[!htbp]
    \centering
    \includegraphics[scale=0.8]{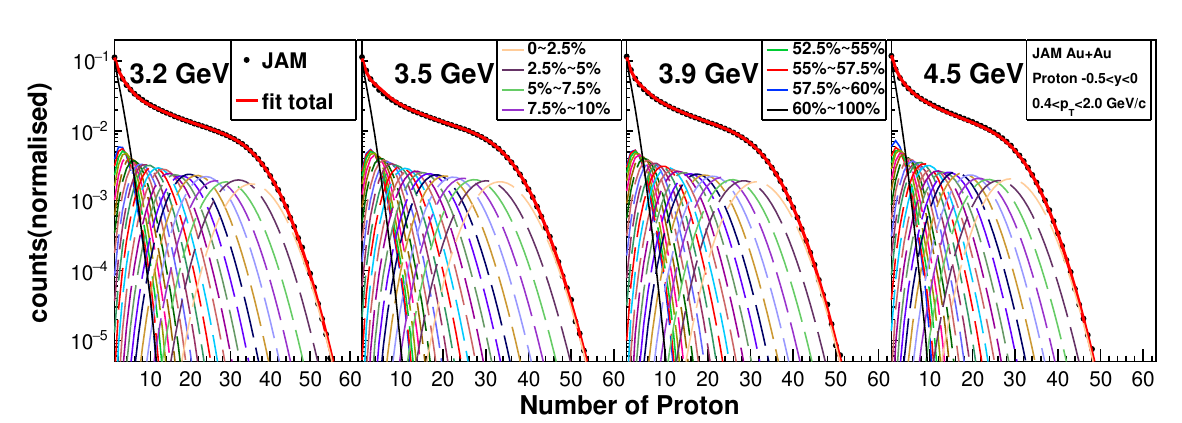}
    \caption{
    Proton number distributions and fitting results at $\sqrt{s_{\rm NN}} = 3.2$--$4.5$~GeV.
    Black solid circles represent JAM model proton number distribution. 
    The 24 colored dashed lines show the sub-distributions for consecutive 2.5\% centrality bins from 0--2.5\% to 57.5--60\%. 
    Black dashed line represents 60--100\% edge events. 
    The red solid line represents the total fitting result, which is the sum of all dashed lines. 
    The proton acceptance window is within $0.4 < p_{T} < 2.0$~GeV/$c$ and $-0.5 < y < 0$.
    }
    \label{fig:PDF_Fit}
\end{figure}
\begin{equation}
    p(x) = \sum_{n=0}^{\infty}c_{n}\frac{d^{n}Z}{dx^{n}}
    \label{eq:GaussianExpanding}
\end{equation}

Here $Z$ represents the standard normal distribution. The coefficients $c_{n}$ are obtained through the inverse Fourier transform of Chebyshev-Hermite polynomials:
\begin{equation}
    c_{n} = \frac{\sqrt{\pi}}{2^{n-1}n!}\int_{-\infty}^{\infty}Z(t)p(t)\,\mathrm{He}_{n}(t)\,dt
    \label{eq:FourierTransform}
\end{equation}

The Chebyshev-Hermite polynomials $\mathrm{He}_{n}$ are defined as:
\begin{equation}
    \mathrm{He}_{n}(x)=(-1)^{n}e^{\frac{x^{2}}{2}}\frac{d^{n}}{dx^{n}}e^{-\frac{x^{2}}{2}}
    \label{eq:ChebyshevHermitePolynomials}
\end{equation}

Therefore, the Edgeworth expansion can be written in the form:
\begin{equation}
    \sigma p(\sigma x) = Z(x)\left\{1+\sum_{s=1}^{\infty}\left[\sigma^{s}\sum_{k_{m}}\mathrm{He}_{s+2r}(x)\prod_{m=1}^{s}\frac{1}{k_{m}!}\left(\frac{S_{m}+2}{m+2}\right)^{k_{m}}\right]\right\}
    \label{eq:EdgeworthExpansion}   
\end{equation}
Here $S_{n}=C_{n}/\sigma^{2n-2}$, and $\sigma$ is the standard deviation of the distribution $p(x)$.

We calculate cumulants and their ratios using CIGAR.
This method employs a differential evolution optimization algorithm 
to determine the optimal high-order moment results, 
ensuring that the sum of each sub-distribution most closely aligns with the simulated distribution. 

\begin{equation}
    \begin{split}
        C_{1}&=a_{c_{1}}n^{4}+b_{c_{1}}n^{3}+c_{c_{1}}n^{2}+d_{c_{1}}n^{1}+e_{c_{1}}\\
        C_{2}&=a_{c_{2}}+b_{c_{2}}C_{1}+c_{c_{2}}C_{1}^{2}\\
        C_{3}&=a_{c_{3}}+b_{c_{3}}C_{1}+c_{c_{3}}C_{1}^{2}\\
        C_{4}/C_{2}&=a_{C_{4}/C_{2}}+b_{C_{4}/C_{2}}C_{1}+c_{C_{4}/C_{2}}C_{1}^{2}
    \end{split}
    \label{eq:cumPara}
\end{equation}

Equation~\ref{eq:cumPara} demonstrates our parameterization strategy, 
in which we describe the cumulants using polynomials. 
This strategy reduces model complexity, 
stabilizes the optimization outcome, and preserves physical significance.

The fitted results are presented in Fig.~\ref{fig:PDF_Fit}. 
Black solid circles represent the proton number distribution from the JAM model. 
The colored dashed lines each correspond to a sub-distribution over a 2.5\% centrality interval, ranging from 0--2.5\% to 57.5--60\%. 
The black dashed line indicates the 60--100\% edge events. 
The red solid line shows the total fitting result, which constitutes the sum of all individual dashed lines. 
The proton acceptance window is defined as $0.4 < p_{T} < 2.0$~GeV/$c$, $-0.5 < y < 0$. 
The corresponding cumulants across all sub-distributions are subsequently calculated based on the fitted parameters of these sub-distributions.

\section{Results and Discussion}\label{analysis}
\begin{figure}[htbp]
    \centering
    \includegraphics[scale=0.5]{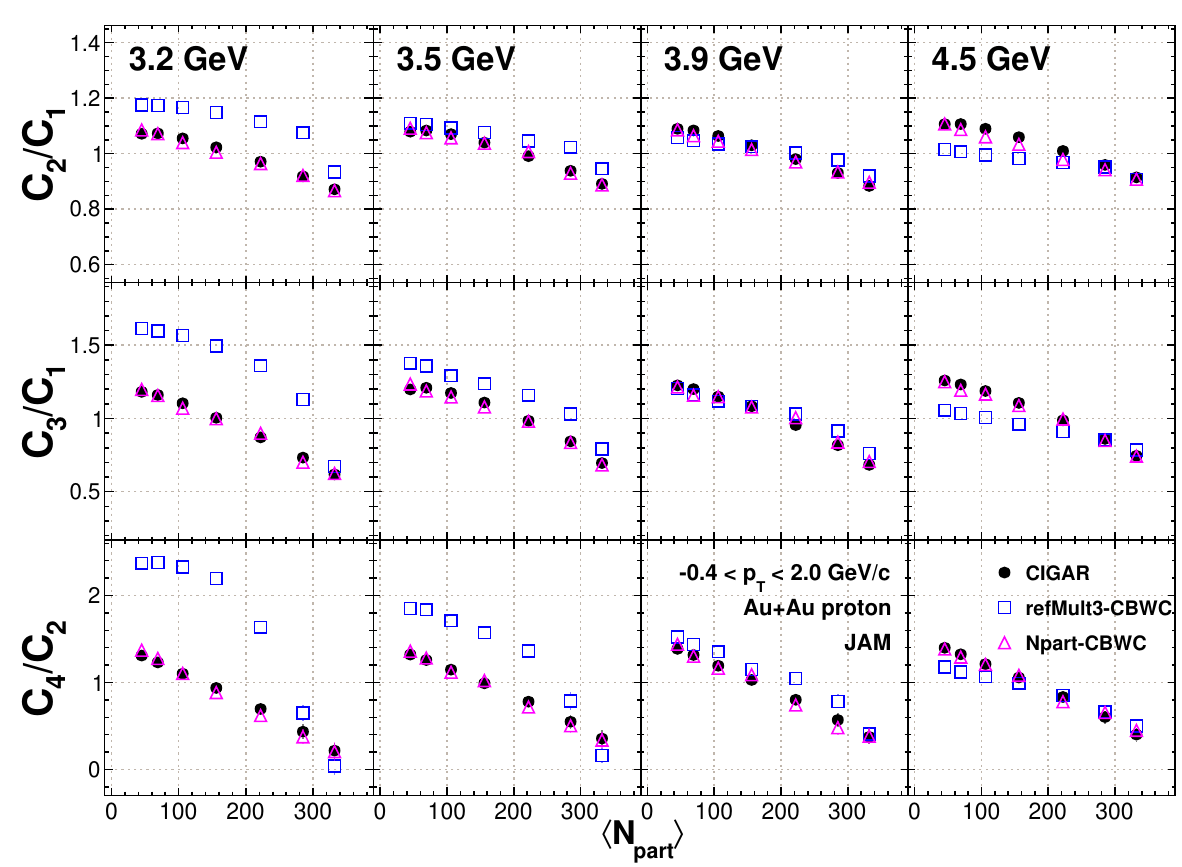}
    \caption{
    Centrality dependence of proton cumulant ratios 
    ($C_{2}/C_{1}$, $C_{3}/C_{1}$, and $C_{4}/C_{2}$) in Au+Au collisions at $\sqrt{s_{\rm NN}} = 3.2$, $3.5$, $3.9$, and $4.5$~GeV  from the JAM model. The black solid circles, blue open squares, and magenta open triangles represent results using CIGAR and CBWC with RefMult3 and $N_{\rm part}$, respectively. The $p_{T}$ range is $0.4 < p_{T} < 2.0$~GeV/$c$.} 
    \label{fig:RatCentDepenMethod}
\end{figure}

In this section, we present the results of proton cumulant, factorial cumulants, and their ratios within the JAM model. We demonstrate in Fig.~\ref{fig:RatCentDepenMethod} the validity of CIGAR method to suppress volume fluctuations at low collision energies, then discuss the effect of spectators on proton cumulants in remaining figures. 

Figure~\ref{fig:RatCentDepenMethod} shows the centrality dependence of $C_{2}/C_{1}$, $C_{3}/C_{1}$ and $C_{4}/C_{2}$ in Au+Au collisions at $\sqrt{s_{\rm NN}} = 3.2$--$4.5$~GeV from the JAM model. The $N_{\rm part}$-based CBWC results (magenta open triangles) serve as a reference baseline with minimal volume fluctuations. We then compare the RefMult3-based CBWC results (blue open squares) and the results using CIGAR method (black solid circles) to this $N_{\rm part}$-based baseline. 

One can see across all four energies, the CIGAR results closely overlap with the $N_{\rm part}$ results, confirming that the CIGAR method effectively eliminates initial volume fluctuations. In contrast, the RefMult3-based CBWC results systematically deviate from the $N_{\rm part}$ baseline, and this deviation is largest at $\sqrt{s_{\rm NN}} = 3.2$~GeV and gradually diminishes toward $4.5$~GeV. This energy-dependent behavior indicates that the CBWC method suffers from reduced centrality resolution at lower collision energies, where the reference multiplicity is smaller, leading to residual volume fluctuations. Furthermore, at each energy, the deviation of RefMult3-based CBWC results from $N_{\rm part}$-based CBWC results gradually increased from central to peripheral, implying that central collisions are least affected volume fluctuation effect. The CIGAR method maintains consistent performance across all energies, establishing itself as a reliable tool for cumulant analysis in the high-baryon-density regime.

\begin{figure}[htbp]
    \centering
    \includegraphics[scale=0.5]{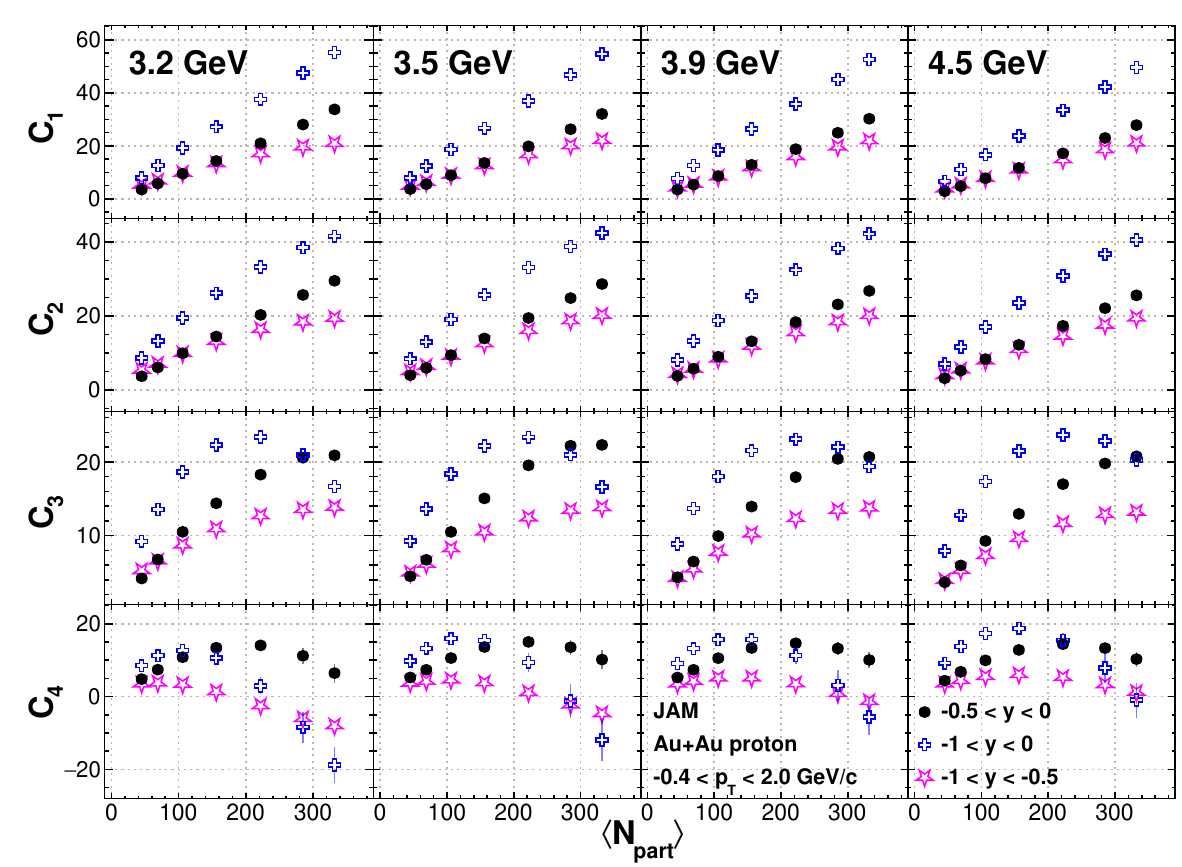}
    \caption{Centrality dependence of proton cumulants up to fourth-order using CIGAR method in Au+Au collisions at $\sqrt{s_{\rm NN}} = 3.2$, $3.5$, $3.9$, and $4.5$~GeV from the JAM model. The black solid circles, blue open crosses, and magenta open stars represent results in rapidity window $-0.5 < y < 0$, $-1 < y < 0$, $-1 < y < -0.5$, respectively. The $p_{T}$ range for analysis is within $0.4<p_{T}<2.0$~GeV/$c$.}
    \label{fig:CumCentDepen}
\end{figure}

Figure~\ref{fig:CumCentDepen} shows the centrality dependence of proton cumulants up to fourth-order using CIGAR method in Au+Au collisions at $\sqrt{s_{\rm NN}} = 3.2$--$4.5$~GeV from the JAM model for three rapidity windows: $-0.5 < y < 0$ (black circles), $-1 < y < 0$ (blue crosses), and $-1 < y < -0.5$ (magenta stars). It is seen that $C_1$ for each rapidity window increases approximately linearly with $N_{\rm part}$, with a slope that decreases at higher collision energies, which can be explained by the weakening of baryon stopping~\cite{PhysRevC.111.024913}.

Notably, $C_4$ exhibits a pronounced non-monotonic behavior: it rises with $N_{\rm part}$, reaches a maximum around $N_{\rm part} \sim 150$--$200$, and then decreases at larger $N_{\rm part}$. This saturation and subsequent suppression in the most central collisions likely arises from baryon number conservation effect, which occurs when the acceptance window constitutes a progressively smaller fraction of the total system volume~\cite{HE2016296}. The wider rapidity window ($-1 < y < 0$) yields systematically larger cumulant values across all orders due to the inclusion of more particles, while the forward rapidity subsample ($-1 < y < -0.5$) shows intermediate values.


\begin{figure}[htbp]
    \centering
    \includegraphics[scale=0.5]{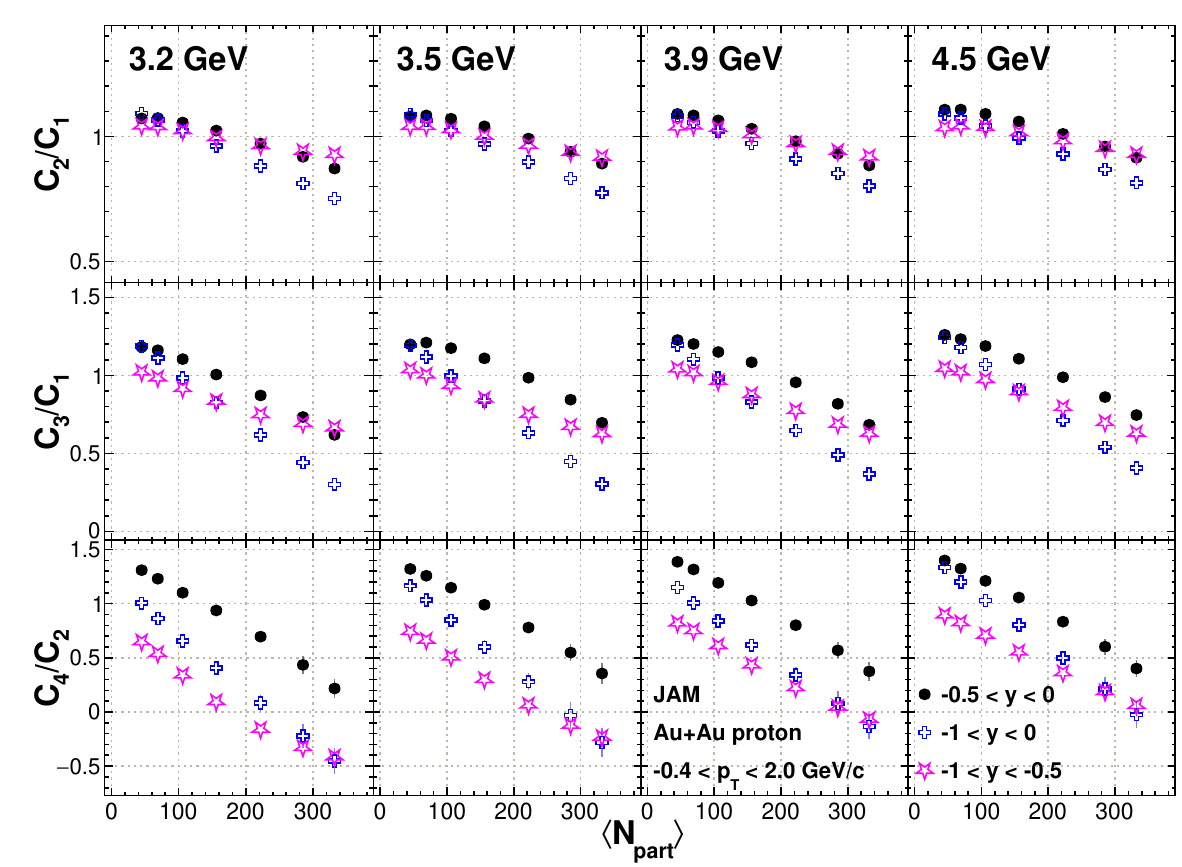}
    \caption{
    Centrality dependence of proton cumulant ratios $C_{2}/C_{1}$, $C_{3}/C_{1}$, and $C_{4}/C_{2}$ in Au+Au collisions at $\sqrt{s_{\rm NN}} = 3.2$, $3.5$, $3.9$, and $4.5$~GeV from the JAM model. The black solid circles, blue open crosses, and magenta open triangles represent results for protons in the rapidity window $-0.5 < y < 0$, $-1 < y < 0$, and $-1 < y < -0.5$, respectively. The $p_{T}$ range for analysis is within $0.4 < p_{T} < 2.0$~GeV/$c$.
    }
    \label{fig:RatCentDepen}
\end{figure}

Figure~\ref{fig:RatCentDepen} shows the centrality dependence of proton cumulant ratios $C_{2}/C_{1}$, $C_{3}/C_{1}$, and $C_{4}/C_{2}$ in Au+Au collisions at $\sqrt{s_{\rm NN}} = 3.2$--$4.5$~GeV for three rapidity windows: $-0.5 < y < 0$ (black circles), $-1 < y < 0$ (blue crosses), and $-1 < y < -0.5$ (magenta stars) from the JAM model. All three ratios exhibit a monotonic decrease from peripheral to central collisions across all energies and rapidity windows. The forward rapidity window ($-1 < y < 0$) yields systematically lower ratio values compared to the mid-rapidity window ($-0.5 < y < 0$) across all centrality bins, with the difference most pronounced for $C_{4}/C_{2}$. The magnitude of these ratio values decreases with increasing collision energy, and the centrality dependence becomes flatter at higher energies.

\begin{figure}[htbp]
    \centering
    \includegraphics[scale=0.5]{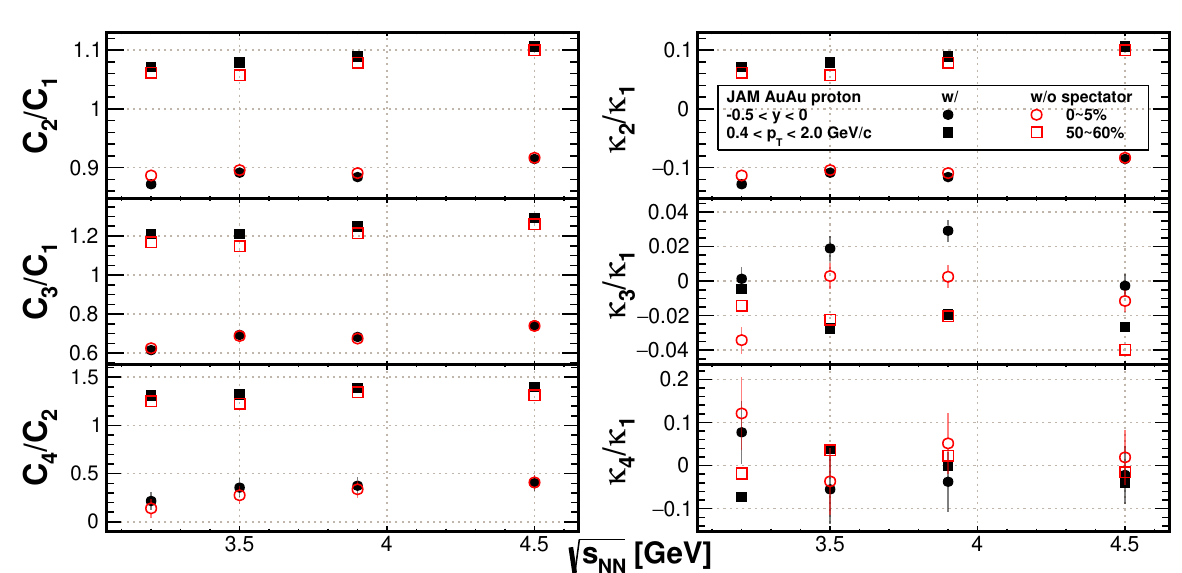}
    \caption{Collision energy dependence of proton cumulant ratios $C_{2}/C_{1}$, $C_{3}/C_{1}$, and $C_{4}/C_{2}$ and factorial cumulant ratios $\kappa_{2}/\kappa_{1}$, $\kappa_{3}/\kappa_{1}$, and $\kappa_{4}/\kappa_{1}$ in Au+Au collisions at $\sqrt{s_{\rm NN}} = 3.2$--$4.5$~GeV from the JAM model. The black solid and open circles represent results in 0--5\% centrality with and without spectator contributions, respectively. The red solid and open squares represent proton results in 50--60\% centrality with and without spectators, respectively. The $p_{T}$ range for analysis is within $0.4 < p_{T} <2.0$~GeV/$c$.}
    \label{fig:EneDepen}
\end{figure}

Figure~\ref{fig:EneDepen} shows the energy dependence of cumulant ratios (left column) and factorial cumulant ratios (right column) in central (0--5\%, black circles) and peripheral (50--60\%, black squares) Au+Au collisions, with red open markers indicating results after spectator removal. For $C_{2}/C_{1}$, spectator removal enhances the ratio in central collisions while reducing it in peripheral collisions, indicating a centrality-dependent spectator effect. For $C_{4}/C_{2}$, spectator exclusion consistently lowers the ratio in both centrality classes, with the reduction most pronounced at 3.2~GeV and diminishing to negligible levels at 4.5~GeV. The factorial cumulant ratios $\kappa_{2}/\kappa_{1}$, $\kappa_{3}/\kappa_{1}$, $\kappa_{4}/\kappa_{1}$ are approximately an order of magnitude smaller than the ordinary cumulant ratios and remain close to zero across all energies. The spectator effect on $\kappa_{4}/\kappa_{1}$ is negligible for both central and peripheral collisions, while $\kappa_{3}/\kappa_{1}$ shows a modest spectator enhancement only in central collisions. 

\begin{figure}[htbp]
    \centering
    \includegraphics[scale=0.6]{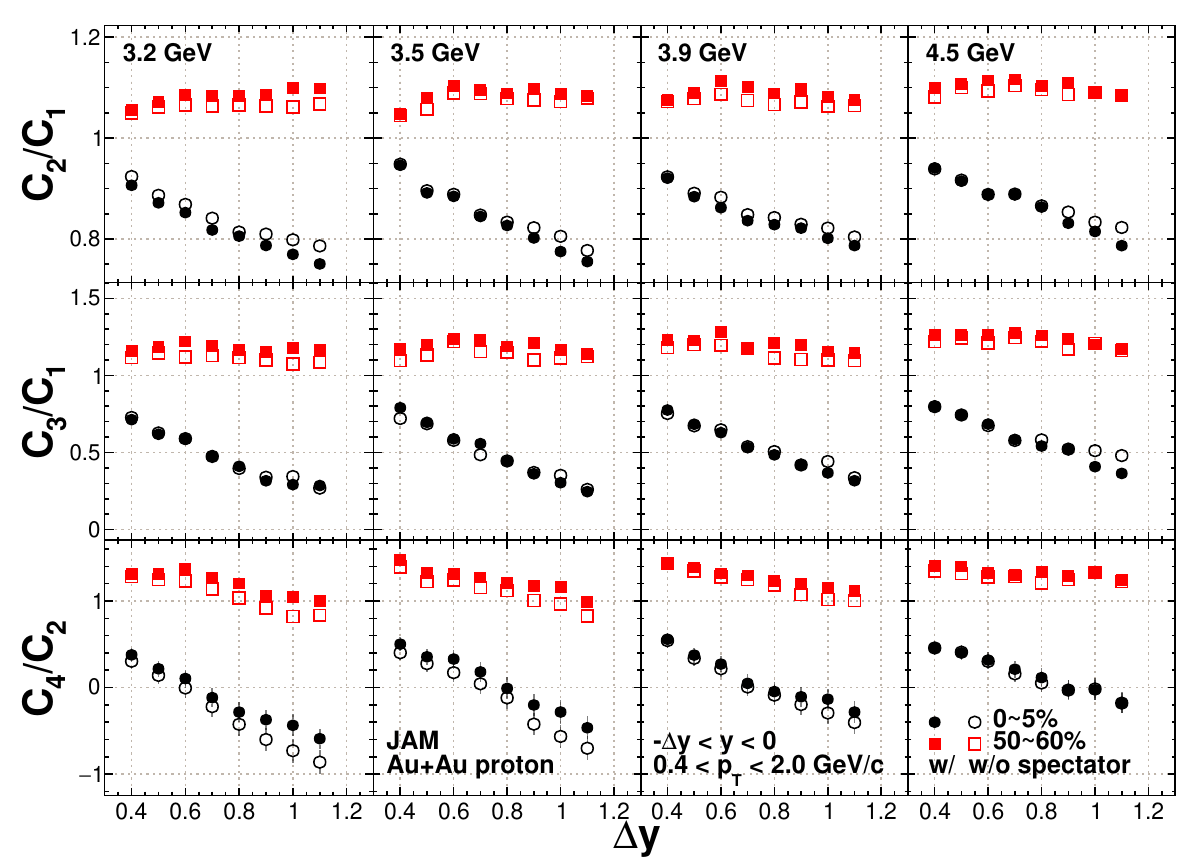}
    \caption{
    Rapidity dependence of proton cumulant ratios $C_{2}/C_{1}$, $C_{3}/C_{1}$, and $C_{4}/C_{2}$ in Au+Au collisions at $\sqrt{s_{\rm NN}} = 3.2$--$4.5$~GeV from the JAM model. The black solid and open circles represent results in 0--5\% centrality bin with and without spectators, respectively. The red solid and open squares represent results in 50--60\% centrality and without spectators, respectively. For all results, the $p_{T}$ range of protons is $0.4 < p_{T} < 2.0$~GeV/$c$ while the rapidity range is $-\Delta y < y < 0$ with $\Delta y$ from 0.4 to 1.1.}
    \label{fig:RapRat}
\end{figure}

We also investigate the spectator effect on the size of rapidity window. Figure~\ref{fig:RapRat} which shows the rapidity dependence of proton $C_{2}/C_{1}$, $C_{3}/C_{1}$, and $C_{4}/C_{2}$ in Au+Au collisions at $\sqrt{s_{\rm NN}} = 3.2$--$4.5$~GeV from the JAM model. In peripheral collisions (red squares), $C_{2}/C_{1}$ and $C_{3}/C_{1}$ remain above unity across all energies and rapidity bins. $C_{2}/C_{1}$ shows a modest increase with $\Delta y$ at 3.2~GeV but flattens at higher energies, while $C_{3}/C_{1}$ exhibits no significant rapidity dependence. $C_{4}/C_{2}$ generally decreases with $\Delta y$ except at 4.5~GeV. In central collisions (black circles), all ratios are substantially below unity and display a strong decreasing trend with $\Delta y$, with the slope becoming less steep at higher collision energies. Spectator removal enhances $C_{2}/C_{1}$ and suppresses $C_{4}/C_{2}$ in central collisions, with the effect most prominent at large $\Delta y$ and low energy. In peripheral collisions, spectator effects are also non-negligible. These observations indicate that spectator contributions predominantly affect higher-order cumulants under conditions of low collision energy and wide rapidity coverage, with their influence diminishing at high energies and in narrow rapidity windows.

\section{Summary}\label{summary}

In summary, we report on proton high-order cumulants, factorial cumulants, and their ratios in Au+Au collisions at $\sqrt{s_{\rm NN}} = 3.2$--$4.5$~GeV from the JAM model. We firstly show that by implementing the CIGAR method, we effectively mitigates the influence of initial volume fluctuations, thereby enabling the extraction of genuine physical signals. The effect of spectator nucleons on cumulant ratios is systematically investigated as a function of both centrality and rapidity window size. Our results show that this effect becomes more pronounced at lower collision energies and larger rapidity window size, which should be taken into careful consideration when comparing with experimental data. This study establishes non-critical baseline measurements for cumulants, factorial cumulants, and their corresponding ratios at high baryon density. Comparisons between our JAM model simulations and the experimental data from the low energy heavy-ion experiments, such as RHIC-STAR fixed-target program at $\sqrt{s_{\rm NN}} = 3.2$--$7.7$~GeV~\cite{Sweger2025} and the future Compressed Baryonic Matter (CBM) experiment at FAIR ($\sqrt{s_{\rm NN}} = 2.4$--$4.9$~GeV)~\cite{CBM2023} may facilitate the searches for the QCD critical point.

\section*{Acknowledgement}
This work is supported in part by the National Key Research and Development Program of China under contract Nos. 2022YFA1604900 and the National Natural Science Foundation of China under Grant No. 12525509 and No. 12447102. Yu Zhang was supported by the Guangxi Natural Science Foundation (No. 2025GXNSFBA069067).

\bibliography{references}
\end{document}